# Electronic Spin: Abstract Mathematical or Real Physical Phenomenon

S K PANDEY,
Department of Mathematics,
Nehru Gram Bharati University,
Kotwa-Jamunipur, Allahabad,
U.P., India.
E-mail:skpandey12@gmail.com

**Abstract**

In the description of electron spin obtained through the conventional Copenhagen interpretation of quantum mechanics, the concrete picture of rotation was replaced by an abstract mathematical representation; visualization or visualisability was entirely lost. The work described here takes a step towards restoring this.

**Keywords**: Dirac equation, spin, size of the electron, rotation, angular velocity, zitterbewegung, geometric algebra.

**PACS Numbers:** 03.65-w, 14.60.cd, 03.65.Fd.

*"The lack of a concrete picture of the spin leaves a grievous gap in our understanding of quantum mechanics".*

H. C. Ohanian [1]

## 1 Introduction

Conventional quantum mechanics does not give a full understanding of what spin is. A physical understanding of spin is still lacking, although spin has become a fundamental feature of scientific and medical technology: consider, for instance, electron spin resonance or magnetic resonance imaging.

T. Koga [2, 3, 4] has given a solution to the Dirac equation and he concludes that the solution represents an anisotropic electron field localized in spacetime. He speculates that the electron field has a form similar to that of a spinning top which has its axis of rotation in a fixed direction. He does not give clear details or mathematical justification for the spin. This led us to study Koga's solution employing the techniques of geometric algebra. Most of this work has already appeared in [11, 12]. The treatment here is quite different and revised.

First, we shall briefly discuss Koga's conjecture in the next section. In the third section we give a refinement of Koga's solution using geometric algebra.

## 2 Koga's Conjecture

The Klein-Gordon equation $\hbar^2 \nabla^2 \phi + m^2 c^4 \phi = 0$ (where $\phi$ is a scalar function) yields the Dirac equation as follows: it can be written as

$$D_0 D_1 \phi = D_1 D_0 \phi = 0.$$

In Koga's notation, for a free electron, we take

$$D_0 = \beta\left(i\hbar \frac{\partial}{\partial t}\right) + \beta\vec{\alpha} \cdot i\hbar c \frac{\partial}{\partial \vec{r}} - mc^2,$$

$$D_1 = \beta\left(i\hbar \frac{\partial}{\partial t}\right) + \beta\vec{\alpha} \cdot i\hbar c \frac{\partial}{\partial \vec{r}} + mc^2.$$

Here $\beta$ is a $4 \times 4$ matrix and $\vec{\alpha}$ is a triple of $4 \times 4$ matrices satisfying well-known anti-commutation relations (see [3], chapters III and V). If $\phi$ is a solution of the Klein-Gordon equation (more precisely, a 4-tuple of scalar solutions) and $|\psi\rangle = D_1 \phi$ then $|\psi\rangle$ satisfies $D_0 |\psi\rangle = 0,$ which is the Dirac equation. Koga takes $\phi_j = a \, \exp\left(\frac{iS}{\hbar}\right) A_j \, \exp\left(i\theta_j\right)$ for $j = 1,2,3,4$ ( $A_j$ and $\theta_j$ are constants and $S$ and $a$ are real scalar fields in spacetime) as four solutions to the Klein-Gordon equation, and

$$|\psi\rangle = \begin{pmatrix} \psi_1 \\ \psi_2 \\ \psi_3 \\ \psi_4 \end{pmatrix} = D_1 \begin{pmatrix} \phi_1 \\ \phi_2 \\ \phi_3 \\ \phi_4 \end{pmatrix}. \qquad (1)$$

We consider an inertial frame in which the electron is at rest. We assume that the origin is, in some sense, the centre of the electron. Then the expressions for $S$ and $a$ can be written as (see [2, 3, 4]): $S = -Ect$, $a = \exp(-\kappa r)/r$, where $\kappa > 0$ and $r = \left| x\hat{i} + y\hat{j} + z\hat{k} \right|$. The energy $cE$ of the electron is defined by $E = -\partial S / \partial(ct)$. We have $E^2 c^2 = m^2 c^4 - \hbar^2 \kappa^2 c^2$. Since $\kappa$ is a constant of nature, so is $E^2 c^2$.

Koga [4] wanted to prove that a free electron has an axis of symmetry and spins around it like a top. For this purpose, Koga considers a rotation of the coordinate system about the z-axis by an angle $\varphi$. He looks at what happens to the four complex components of the solution: $\psi_1, \psi_2, \psi_3, \psi_4$.

For instance, the expression for $\psi_1$ has four terms, each containing only one constants $\theta_j$ in its argument. After the rotation, he shows that in the case of $\psi_1$, $\theta_4$ gets replaced by $\theta_4 - \varphi$. But at this point, he asserts that since the angle $\theta_j$ is arbitrary, we can simply ignore the effect of $\varphi$. We consider this unsatisfactory.

## 3 Refinement of Koga's Solution to the Dirac Equation Using Geometric Algebra

The geometric algebra of the Minkowski spacetime is called spacetime algebra (STA). It is the 16 dimensional real algebra generated by $\gamma_0, \gamma_1, \gamma_2, \gamma_3$ with relations

$$\gamma_0^2 = 1, \gamma_0 \cdot \gamma_m = 0, \gamma_m \cdot \gamma_k = 0 \text{ if } m \neq k \text{ and } = -1 \text{ if } m = k, (m, k = 1, 2, 3).$$

The members of STA are called multivectors. A multivector is a sum of terms, each term being a real number times a product of $\gamma_i$'s. A term is called even if the number of $\gamma_i$ factors is even. An even multivector is a sum of even terms. These form an 8-dimensional subalgebra of STA, called the even subalgebra. For more details, one may refer to [5, 6]. The STA version of the usual Dirac equation is called now the Dirac-Hestenes equation. This was first introduced by Hestenes [5]. It is given by [6]:

$$\hbar \nabla \psi I \sigma_3 = mc \psi \gamma_0 \qquad (2)$$

Here $\nabla \psi = \gamma^0 \dfrac{\partial \psi}{\partial(ct)} + \gamma^1 \dfrac{\partial \psi}{\partial x} + \gamma^2 \dfrac{\partial \psi}{\partial y} + \gamma^3 \dfrac{\partial \psi}{\partial z}$, $I = \gamma_0 \gamma_1 \gamma_2 \gamma_3 = \sigma_1 \sigma_2 \sigma_3$ is called the pseudoscalar of STA and $\sigma_k = \gamma_k \gamma_0$. It may be noted that $\gamma^\mu, (\mu = 0,1,2,3)$ are the reciprocal frame of vectors with respect to the frame $\gamma_\mu$ satisfying $\gamma^0 = \gamma_0$ and $\gamma^k = -\gamma_k, (k = 1,2,3)$.

We can rewrite the usual Dirac equation $D_0 |\psi\rangle = 0$ as

$$i\hbar\left(\hat{\gamma}^0\frac{\partial}{\partial t}+c\hat{\gamma}^1\frac{\partial}{\partial x}+c\hat{\gamma}^2\frac{\partial}{\partial y}+c\hat{\gamma}^3\frac{\partial}{\partial z}\right)|\psi\rangle=mc^2|\psi\rangle$$

and equation (1) can be put in the following form

$$|\psi\rangle=\left(i\hbar\hat{\gamma}^0\frac{\partial}{\partial t}+i\hbar c\hat{\gamma}^1\frac{\partial}{\partial x}+i\hbar c\hat{\gamma}^2\frac{\partial}{\partial y}+i\hbar c\hat{\gamma}^3\frac{\partial}{\partial z}+mc^2\right)ae^{\left(\frac{iS}{\hbar}\right)}A_j e^{(i\theta_j)}$$

After some calculations this becomes

$$|\psi\rangle=a\exp\left(\frac{iS}{\hbar}\right)\{Ec\hat{\gamma}^0+mc^2-i\hbar c\hat{\gamma}^1 R_x-i\hbar c\hat{\gamma}^2 R_y-i\hbar c\hat{\gamma}^3 R_z\}\begin{pmatrix}A_1 e^{(i\theta_1)}\\ A_2 e^{(i\theta_2)}\\ A_3 e^{(i\theta_3)}\\ A_4 e^{(i\theta_4)}\end{pmatrix} \quad (3)$$

Here $\hat{\gamma}^\mu$ are the standard Dirac-Pauli matrices. We have taken $\beta=\hat{\gamma}^0, \beta\alpha_x=\hat{\gamma}^1$, $\beta\alpha_y=\hat{\gamma}^2, \beta\alpha_z=\hat{\gamma}^3$ and $R_x=x\left(\frac{1}{|\vec{r}|^2}+\frac{\kappa}{|\vec{r}|}\right)$ etc.

Hestenes [7] has shown that an arbitrary spinor $|\psi\rangle$ satisfying the Dirac equation can always be transformed into an even multivector $\psi:|\psi\rangle\leftrightarrow\psi u$. We eliminate complex imaginary unit $i$ from the above solution. This is achieved by putting

$$\begin{pmatrix}A_1 e^{(i\theta_1)}\\ A_2 e^{(i\theta_2)}\\ A_3 e^{(i\theta_3)}\\ A_4 e^{(i\theta_4)}\end{pmatrix}=\begin{pmatrix}1\\0\\0\\0\end{pmatrix}=u,$$

$$i\hat{\gamma}^1 u=\hat{\gamma}^2\hat{\gamma}^0 u,\ -i\hat{\gamma}^2 u=\hat{\gamma}^1\hat{\gamma}^0 u,\ iu=\hat{\gamma}^2\hat{\gamma}^1 u\text{ etc.}$$

We also use

$$\hat{\gamma}^0 u=u,\ \hat{\gamma}^0=\hat{\gamma}_0,\ \hat{\gamma}^k=-\hat{\gamma}_k,(k=1,2,3).$$

Using above relations equation (3) reduces to

$$|\psi\rangle=a\{(Ec+mc^2)+(\hbar cR_x\hat{\gamma}_2\hat{\gamma}_0-\hbar cR_y\hat{\gamma}_1\hat{\gamma}_0+\hbar cR_z\hat{\gamma}_0\hat{\gamma}_1\hat{\gamma}_2\hat{\gamma}_3)\}\left(\cos\frac{S}{\hbar}+\hat{\gamma}_2\hat{\gamma}_1\sin\frac{S}{\hbar}\right)u$$

Following Hestenes we can treat the matrices $\hat{\gamma}_0,\hat{\gamma}_1,\hat{\gamma}_2,\hat{\gamma}_3$ as the generators (vectors) $\gamma_0,\gamma_1,\gamma_2,\gamma_3$ of spacetime algebra and then the coefficient of $u$ gives the geometric algebra version of (3) as

$$\psi=a\{(Ec+mc^2)+(\hbar cR_x\gamma_2\gamma_0-\hbar cR_y\gamma_1\gamma_0+\hbar cR_z I)\}\exp\left(I\sigma_3\frac{S}{\hbar}\right) \quad (4)$$

Here $\vec{R}=(R_x,R_y,R_z)=\vec{r}\left(\frac{1}{r^2}+\frac{\kappa}{r}\right)$ and $\vec{r}=x\gamma_1+y\gamma_2+z\gamma_3=(x,y,z)$.

Equation (4) is a solution to the Dirac-Hestenes equation (2). This solution can directly be derived from the Dirac-Hestenes equation without following the method described here. For derivation and further details, one may refer to [11, 12]. After some calculations equation (4) can be seen to be equivalent to the solution given in [11, 12].

Like Koga, we interpret the solution as suggesting that the electron is a localized field, with the value of $\psi$ at each point giving the properties of the electron at that point, at least in principle.

The above solution has two terms: the first term is $a(Ec + mc^2)\exp\left(I\sigma_3 \dfrac{S}{\hbar}\right)$. This satisfies the Klein-Gordon equation $\hbar^2 \nabla^2 \varphi + m^2 c^2 \varphi = 0$, where $\varphi$ is an even multivector and it stands for a localized field without spin.

The second term can be written as a sum of two terms as under:

$$a(\hbar c R_x \gamma_2 \gamma_0 - \hbar c R_y \gamma_1 \gamma_0 + \hbar c R_z I)\exp\left(I\sigma_3 \frac{S}{\hbar}\right)$$

$$= a(\hbar c R_x \gamma_1 \gamma_0 + \hbar c R_y \gamma_2 \gamma_0 + \hbar c R_z \gamma_3 \gamma_0)\exp\left(\frac{S}{\hbar} + \frac{\pi}{2}\right)I\sigma_3$$

$$= \{ a(\hbar c R_x \sigma_1 + \hbar c R_y \sigma_2)\exp\left(\frac{S}{\hbar} + \frac{\pi}{2}\right)I\sigma_3 + a\hbar c R_z \sigma_3 \}$$

$$+ \{ a\hbar c R_z \sigma_3 \left(\exp\left(\frac{S}{\hbar} + \frac{\pi}{2}\right)I\sigma_3 - 1\right) \} \quad (4)$$

The first term on the right equals

$$a\hbar c \left\{ e^{-\frac{(S/\hbar + \pi/2)}{2}I\sigma_3} (R_x \sigma_1 + R_y \sigma_2 + R_z \sigma_3) e^{\frac{(S/\hbar + \pi/2)}{2}I\sigma_3} \right\}$$

$$= a\hbar c \left(\frac{1}{r^2} + \frac{\kappa}{r}\right) \left\{ e^{-\frac{(S/\hbar + \pi/2)}{2}I\sigma_3} (x\sigma_1 + y\sigma_2 + z\sigma_3) e^{\frac{(S/\hbar + \pi/2)}{2}I\sigma_3} \right\}$$

which exhibits a spinning field with a constant angular velocity $\omega = \dfrac{-Ec}{\hbar}$ about $\sigma_3$ axis (i.e. in the $\sigma_1 \sigma_2$ plane). We can estimate the spin angular velocity of the electron. The angular velocity is roughly found to be of the order of $10^{21}$ radians per second [9]. This theory enables us to put a bound on the size of the electron field. Considering that speeds greater than $c$ do not occur we have $r < c/\omega$. Here $r$ is the distance from any point in the electron to the axis of rotation. Roughly, this theory sets an upper bound on the size of a free electron at rest at less than $10^{-14}$ meter.

The second term on the right of (4) equals to

$$a\hbar c\left(\frac{1}{r^2}+\frac{\kappa}{r}\right)\left\{z\sigma_3\left(\exp\left(\frac{S}{\hbar}+\frac{\pi}{2}\right)I\sigma_3 - 1\right)\right\}$$

At each point $x\sigma_1 + y\sigma_2 + z\sigma_3$, the value of the expression given in the braces is independent of $x$ and $y$ and proportional to $z$. It represents an oscillatory motion that gets larger with $|z|$. This is modulated by a scalar factor which makes it a localized field which vanishes at infinity. This motion seems similar (or at least vaguely analogous) to Schrodinger's Zitterbewegung. However, there is no interference between the states, in fact there are no states in Koga's theory, only a pointwise description of the electron field. The Zitterbewegung described here has nothing to do with negative energy and it is in the direction of the spin axis. The Zitter frequency is roughly of the order of $10^{20}$ Hertz.

Finally, it may be noted that by considering both the Dirac equations given by the Klein-Gordon equation we get four solutions, two for each Dirac equation and thus we get all four combinations of energy and spin. Koga's view should be noted ([3], sec. 3.3): the superposition principle is not valid, the motion of an individual electron is always casual and continuous and there is no question of its jumping into a negative energy state. The idea of a vacuum filled with negative energy electrons, and pairs of virtual electrons and positrons, should be abandoned.

## 4 Summary and Concluding Remarks

The solution to the Dirac equation using geometric algebra and its implication is not what we expected. We thought we would just get a spinning field as Koga stated. The first surprise was the Klein-Gordon term which, with hindsight, we should have expected. Then the rest of the solution also did not represent a spinning field. Almost miraculously, it turned out that it can be broken up into a spinning term and another term showing a one-dimensional oscillation modulated by a scalar factor, which makes it a localized field.

The last two terms (as mentioned above) can be considered corrections to the Klein-Gordon term, which stands for a particle (or field) without spin. This seems roughly parallel to the history of the subject: Schrodinger first considered (and rejected) the Klein-Gordon equation, and spin was discovered a bit latter. Zitterbewegung or shudder came still latter.

The spin and Zitterbewegung have been discussed by many authors but it seems unlikely to us that there is anything similar to this work in the literature. The solution described here is not similar to any well known solution. For example, Doran and Lasenby [6] only talk about plane waves. So does Baylis ([8], Section 19.6). Hestenes has done a great deal of work on the Dirac theory over several decades using geometric algebra. He considers a point electron moving along a helix in Minkowski space, corresponding to uniform circular motion in 3-space. Hestenes [9, 10] does not give any similar solution. He only considers plane wave solutions.

It seems that Koga's solution to the Dirac equation and its refinement using geometric algebra possibly hold the key to understand what is electronic spin: may be the mysterious phenomenon that various observations hint at is really a spinning and shuddering electron!